\documentclass[a4paper,fleqn,usenatbib]{pack/mnras}
\usepackage[T1]{fontenc}
\usepackage{ae,aecompl}
\usepackage{graphicx}	
\usepackage{amsmath}	
\usepackage{amssymb}	
\usepackage{courier}
\usepackage{pack/multirow}
\usepackage{pack/soul}
\usepackage{pack/float}
\usepackage{pack/gensymb}
\usepackage{pack/color}
\usepackage{pack/threeparttable}
\title[GMASS 0953 - ALMA]{ALMA view of a massive spheroid progenitor: a compact rotating core of molecular gas in an AGN host at z=2.226}
\author[M. Talia et al.]{M.Talia,$^{1,2}$\thanks{E-mail: margherita.talia2@unibo.it}
F.Pozzi,$^{1,2}$ L.Vallini,$^{3}$ A.Cimatti,$^{1,4}$ P.Cassata,$^{5}$ F.Fraternali,$^{1,6}$ M.Brusa,$^{1,2}$  
\newauthor E.Daddi,$^{7}$ I.Delvecchio,$^{8}$ E.Ibar,$^{5}$ E.Liuzzo,$^{9}$ C.Vignali,$^{1}$ M.Massardi,$^{9}$ G.Zamorani,$^{2}$   
\newauthor C.Gruppioni,$^{2}$ A.Renzini,$^{10}$ M.Mignoli,$^{2}$ L.Pozzetti,$^{2}$ G.Rodighiero$^{11}$\\
$^{1}$Dipartimento di Fisica e Astronomia, Universit\`a di Bologna, Via Gobetti 93/2, I-40129, Bologna, Italy\\
$^{2}$INAF- Osservatorio Astronomico di Bologna, Via Gobetti 93/3, I-40129, Bologna, Italy\\ 
$^{3}$Nordita, KTH Royal Institute of Technology and Stockholm University, Roslagstullsbacken 23, SE-10691 Stockholm, Sweden\\
$^{4}$INAF - Osservatorio Astrofisico di Arcetri, Largo E. Fermi 5, I-50125, Firenze, Italy \\
$^{5}$Instituto de Fisica y Astronom\'ia, Facultad de Ciencias, Universidad de Valpara\'iso, Gran Breta\~na 1111, Playa Ancha, Valpara\'iso, Chile\\
$^{6}$University of Groningen, Kapteyn Astronomical Institute, Postbus 800, NL-9700 AV Groningen, the Netherlands\\
$^{7}$CEA, IRFU, DAp, AIM, Universit\'e Paris-Saclay, Universit\'e Paris Diderot, Sorbonne Paris Cit\'e, CNRS, F-91191 Gif-sur-Yvette, France\\
$^{8}$Department of Physics, Faculty of Science, University of Zagreb, Bijeni\v{c}ka cesta 32, 10000 Zagreb, Croatia\\
$^{9}$INAF, Istituto di Radioastronomia - Italian ARC, Via Piero Gobetti 101, I-40129 Bologna, Italy\\
$^{10}$INAF-Osservatorio Astronomico di Padova, Vicolo dell'Osservatorio 2, I-35122 Padova, Italy\\
$^{11}$Dipartimento di Fisica e Astronomia G. Galilei, Universit\`a di Padova, Vicolo dell'Osservatorio 3, I-35122, Italy
}
\date{Accepted XXX. Received YYY; in original form ZZZ}
\pubyear{2017}
\begin{document}
\label{firstpage}
\pagerange{\pageref{firstpage}--\pageref{lastpage}}
\maketitle
\begin{abstract}
We present ALMA observations at 107.291 GHz (band 3) and 214.532 GHz (band 6) of GMASS 0953, a star-forming galaxy at $z$=2.226 hosting an obscured AGN that has been proposed as a progenitor of compact quiescent galaxies (QG).
We measure for the first time the size of the dust and molecular gas emission of GMASS 0953 that we find to be extremely compact ($\sim$1 kpc). 
This result, coupled with a very high ISM density ($n\sim10^{5.5}$ cm$^{-3}$), a low gas mass fraction ($\sim$0.2) and a short gas depletion timescale ($\sim$150 Myr) imply that GMASS 0953 is experiencing an episode of intense star-formation in its central region that will rapidly exhaust its gas reservoirs, likely aided by AGN-induced feedback, confirming its fate as a compact QG. 
Kinematic analysis of the CO(6-5) line shows evidence of rapidly-rotating gas ($V_{rot}$=320$^{+92}_{-53}$ km s$^{-1}$), as observed also in a handful of similar sources at the same redshift. 
On-going quenching mechanisms could either destroy the rotation or leave it intact leading the galaxy to evolve into a rotating QG. 
 
\end{abstract}
\begin{keywords}
galaxies: high-redshift -- galaxies:evolution -- ISM:kinematics and dynamics -- galaxies:active 
\end{keywords}
%
%
%
\section{Introduction}\label{sec:Introduction}
Massive star-forming galaxies (SFGs) with centrally concentrated luminosity profiles at $z$>2 have been recently suggested by different authors  to be the direct progenitors of compact quiescent galaxies (cQGs) at $z$=1.5-3 \citep[e.g.][]{wuyts2011, whitaker2012a, barro2013, vandokkum2015, toft2007, cassata2011, vanderwel2014}.
Several theories have been proposed to achieve the high stellar densities observed in cQGs, including gas-rich mergers and/or disk instabilities \citep[e.g.][]{tacconi2008, zolotov2015, tacchella2016}, or in-situ inside-out growth \citep[e.g.][]{wellons2015, lilly2016}.

Most scenarios predict the formation of a compact SFG (cSFG) as the last stage before quenching the star formation. 
Observationally, cSFGs candidates have been identified as being dense, compact, and dusty \citep{barro2013, nelson2014, vandokkum2015}.
The kinematics of the H$\alpha$ emission line suggests that cSFGs have rotating disk of ionized gas slightly larger or comparable to the stellar distribution \citep[][]{vandokkum2015, wisnioski2017}.
From the comparison between the dynamical and stellar masses it is inferred that these galaxies must have low gas mass fractions and short gas depletion timescales, as would be expected if they were soon to terminate their star formation.

A key element in the understanding of quenching mechanisms is a direct measurement of the size and mass content of the cold gas reservoirs that provide the fuel for star formation, but until now there have been only few of such measurements \citep[e.g.][]{barro2016, spilker2016, popping2017, tadaki2017b}.
cSFGs seem also to show a higher AGN incidence than the overall galaxy population at a fixed stellar mass \citep{kocevski2017, wisnioski2017}, suggesting that AGN activity might play a role in quenching, possibly through feedback provided by large-scale outflows \citep[e.g.][]{gilli2014, genzel2014, brusa2015b}. 

In order to investigate the gas properties of the progenitors of cQGs, in this letter we present ALMA spatially resolved observations of the dust continuum and CO lines emission of GMASS 0953, a heavily obscured AGN host selected from the GMASS sample \citep{kurk2013}.
We adopt a cosmology with $H_{0}{=}70$ km s$^{-1}$ Mpc$^{-1}$, $\Omega_{m}{=}0.3$, $\Omega_{\Lambda}{=}0.7$ and assume a \citet{chabrier2003} IMF.
%
%
%
    \begin{figure}
	\centering
	\includegraphics[scale=0.35]{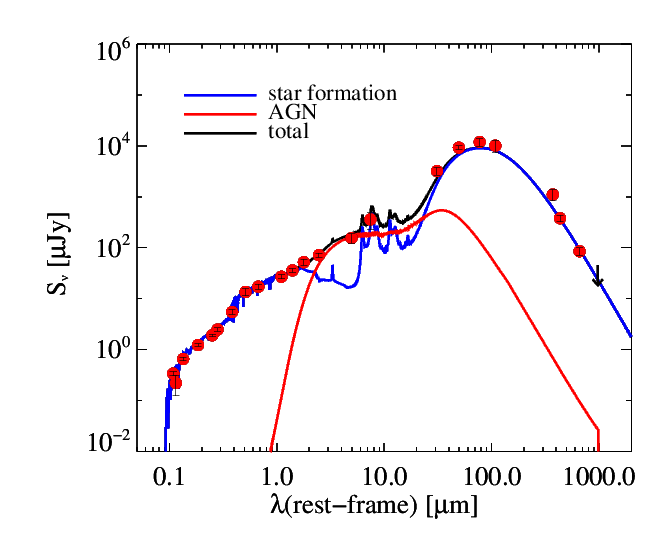}
	\caption{SED of GMASS 0953. 
The black line is the total best fit model, while blue and red curves indicate respectively the star formation and AGN contributions. 
Red dots mark the observed photometry from: MUSIC \citep{grazian2006}, SPITZER/MIPS \citep{magnelli2011}, \emph{Herschel}/PACS \citep{magnelli2013} and SPIRE \citep{roseboom2010}, ALMA at 1.2mm \citep{ueda2018}, 1.4 mm (this work, Sec. \ref{sec:data}), and 2.1 mm \citep{popping2017}. The black arrow represents the 5$\sigma$ upper limit on the ALMA band 3 continuum obtained from the combined map of our data with those by \citet{popping2017}.}
	\label{ivanplot}
	\end{figure}
%
%
%
    \begin{figure}
	\centering
	\includegraphics[scale=0.5, trim= 0mm 0mm 140mm 0mm, clip=true]{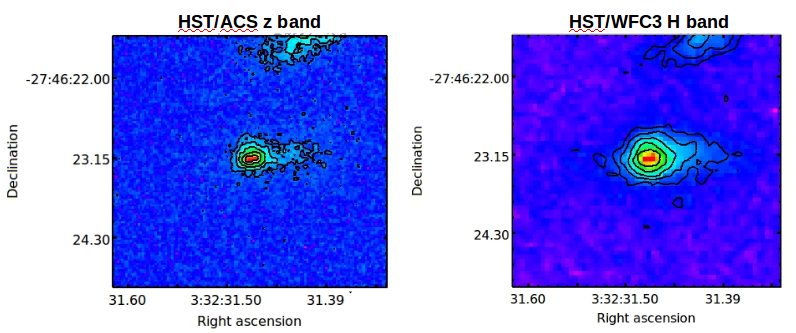}
	\includegraphics[scale=0.5, trim= 140mm 0mm 0mm 0mm, clip=true]{Figure/hst.png}
	\caption{HST/ACS z-band \citep[top,][]{giavalisco2004} and CANDELS HST/WFC3 H-band \citep[bottom,][]{grogin2011}. The lower contours are at 3$\sigma$ level. The source to the north of our target is a foreground galaxy.}
	\label{hst}
	\end{figure}
%
%
%
\section{GMASS 0953}\label{sec:gmass0953}
GMASS 0953\footnote[1]{a.k.a. K20-ID5, GS3-19791, 3D-HST GS30274 \citep[e.g.][]{daddi2004b, forsterschreiber2009, popping2017}.} (R.A. 03:32:31.48, Dec. -27:46:23.40) is a SFG at z$_{CO}$=2.2256.
It is detected in the 7Ms CDF-S X-ray maps \citep{luo2017} and hosts a heavily obscured ($N_{\rm H}$>10$^{24}$ cm$^{-2}$; Dalla Mura et al. in prep.) AGN with a rest-frame intrinsic luminosity (i.e. corrected for the obscuration) $L_{\rm 2-10keV}{\sim}$6.0$\times$10$^{44}$ erg/s.
The target shows marginally extended emission in the 1.4 GHz VLA radio continuum maps \citep{miller2013}. The monochromatic 1.4 GHz luminosity $L_{\rm 1.4GHz}$ = 10$^{24.84}$ W Hz$^{-1}$ is consistent with radio emission predominantly arising from an AGN \citep{bonzini2012, bonzini2013}.
Despite the clear presence of the AGN, the rest-frame UV spectrum remarkably does not show high-ionization emission lines (e.g. CIV$\lambda$1550, SiIV$\lambda$1400), likely because of the large obscuration of the nucleus \citep{cimatti2013}.
Optical lines ratios are consistent with a type II AGN, though shocks have also been proposed as an ionization mechanism \citep{vandokkum2005}, supported by evidence of large-scale outflows in multiple gas phases \citep[][Loiacono et al. in prep.]{cimatti2013, forsterschreiber2014, genzel2014}.

Following \citet{delvecchio2014}, we performed a multi-component SED fitting to the available broadband photometry with the SED3FIT code \citep{berta2013}, that combines \citet{bruzual2003} stellar libraries, \citet{dacunha2008} IR-dust libraries, and \citet{feltre2012} torus+disc models.
The full SED is shown in Fig. \ref{ivanplot}.
We derive a the stellar mass $M_{\star}$=(1.15$\pm$0.1)$\times$10$^{11}$ $M_{\odot}$ and $SFR_{\rm IR}$=214$\pm$20 $M_{\odot}$ yr$^{-1}$, the latter assuming the \citet{kennicutt1998} relation (scaled to a \citet{chabrier2003} IMF) between rest-frame 8-1000$\mu$m $L_{\rm IR}$, corrected for the AGN contribution, and SFR.
These values would place GMASS 0953 on the SFR-mass main sequence \citep[MS; e.g.][]{rodighiero2011}. 
A fit of the FIR points ($\lambda$>24$\mu$m) to a greybody gives values of $M_{\rm dust}$=(2.6$\pm$0.5)$\times$10$^{8}$ $M_{\odot}$ and $T_{\rm dust}$=38$\pm$2 K, consistent with \citet{popping2017}.
%
%
%
    \begin{figure*}
	\centering
	\includegraphics[scale=0.34, trim= 95mm -8mm 110mm 25mm, clip=true]{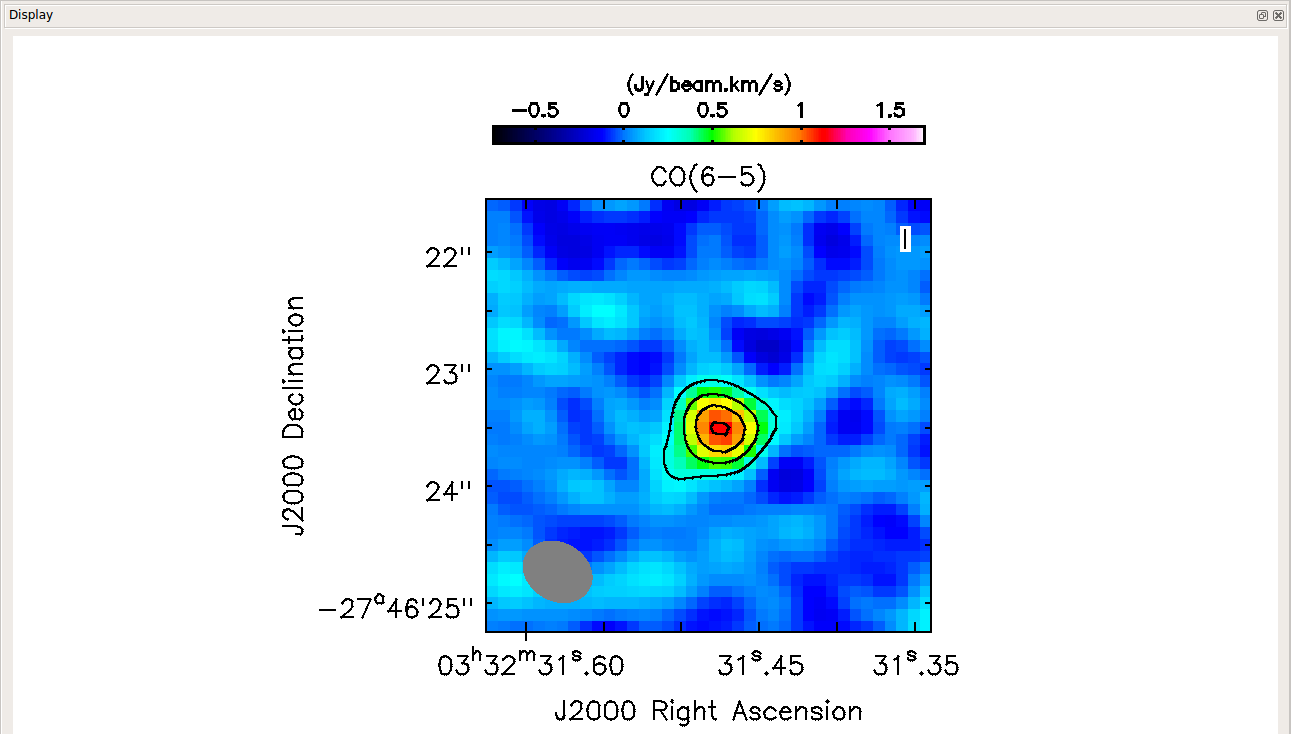}
	\includegraphics[scale=0.368, trim=105mm 10mm 110mm 20mm, clip=true]{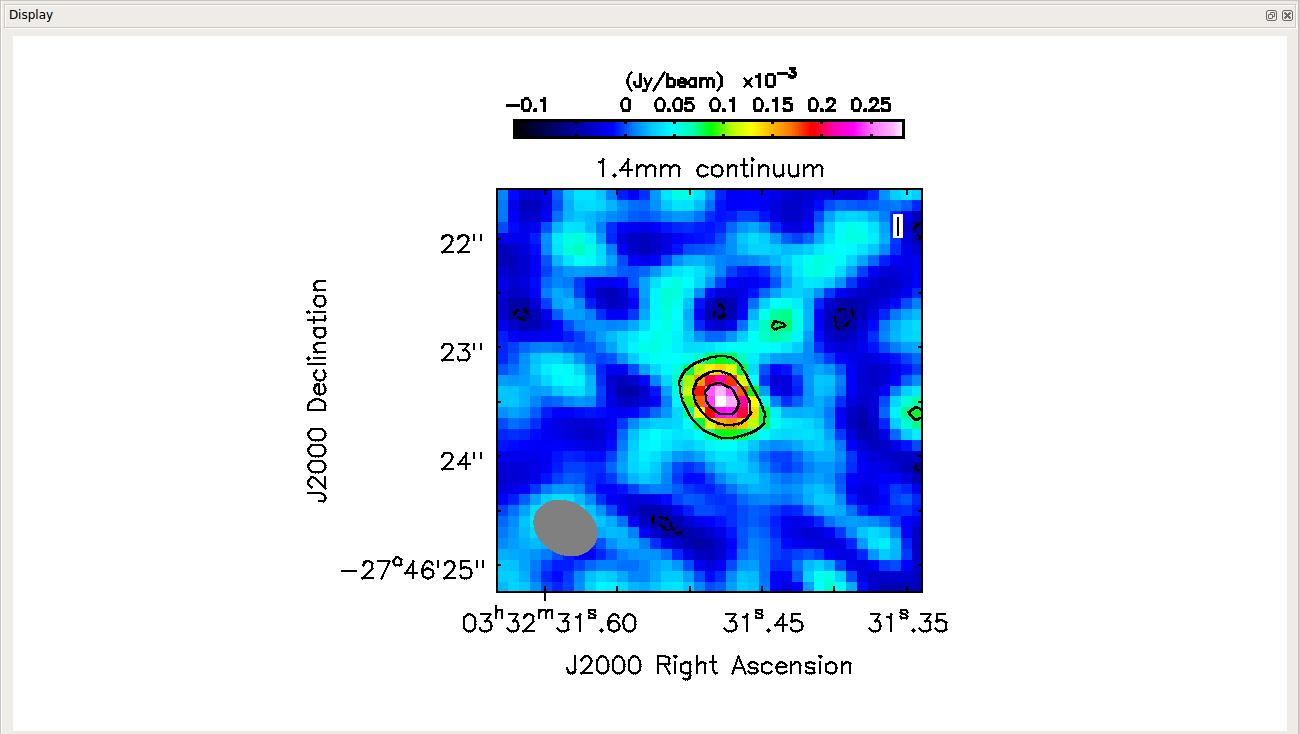}
	\includegraphics[scale=0.39, trim=110mm 10mm 110mm 20mm, clip=true]{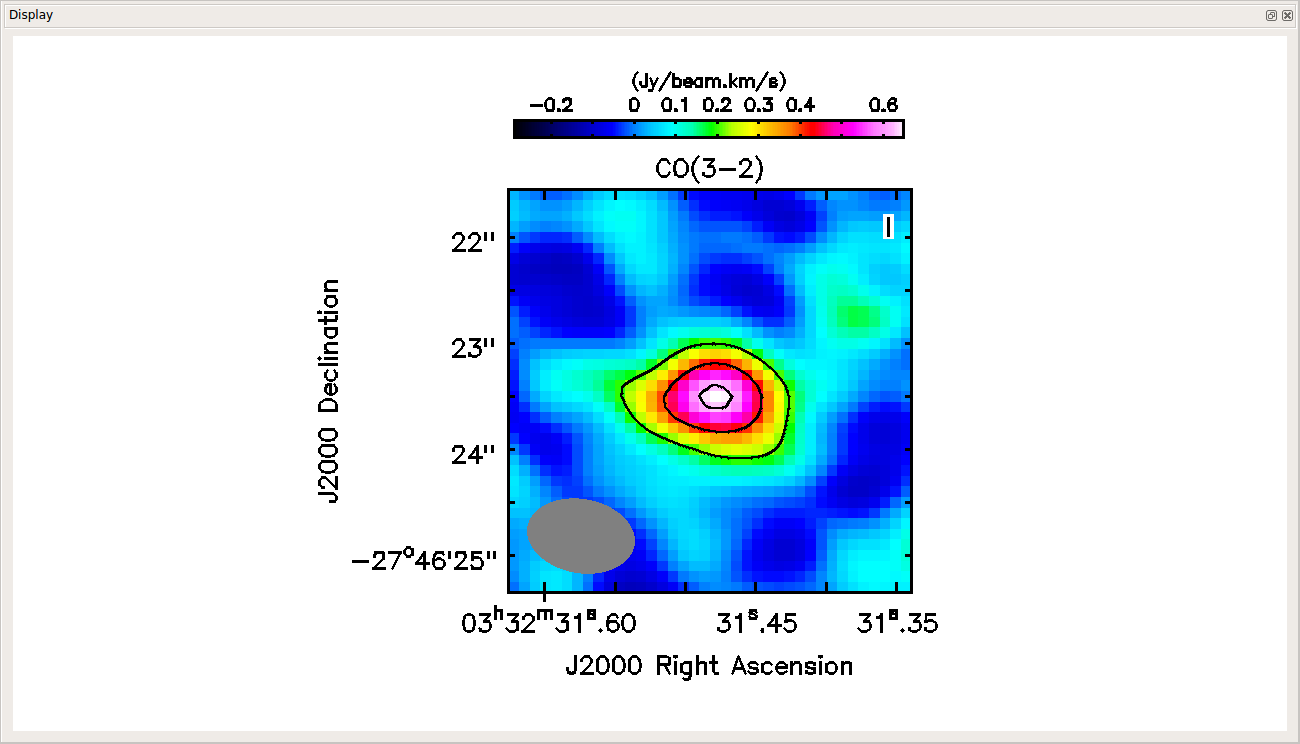}
	\caption{\emph{From left to right:} ALMA 1.4mm continuum map (band 6), \emph{moment 0} map of the CO(3-2) line, \emph{moment 0} map of the CO(6-5) line. The beam size is also shown in grey. In all images the lower continuous contours are at 3$\sigma$ level. The scales are the same as in Fig. \ref{hst}, but no astrometric correction was applied (see Sec. \ref{sec:data}).}
	\label{alma}
	\end{figure*}

HST images (Fig. \ref{hst}; top: ACS/z-band; bottom: WFC3/H-band)) show a compact morphology \citep[$r_{\rm e,Hband}$=2.5 kpc,][]{vanderwel2014} with a low-surface brightness tail to the west of the core that has been interpreted as either a merger remnant \citep[][]{vandokkum2015} or a faint disk \citep[][]{wisnioski2017}.
%
%
%
\section{ALMA observations}\label{sec:data}
ALMA observations were carried out in band 3 and 6, during Cycle 3 project 2015.1.01379.S (PI: P. Cassata) for a total integration time on source of 32 mins and 1.3 hrs respectively, and an angular resolution of 0.6$\arcsec$.
The precipitable water vapour during the observations was between 1.4 mm and 3.1 mm.
We centred one spectral window of bandwidth 1.875 GHz covering 3840 channels at 107.291 GHz and 214.532 GHz respectively in band 3 and 6 to target CO(3-2) and CO(6-5) lines, and placed  in each band, on line-free regions, other two spectral windows of bandwidth 1.875 GHz covering 960 channels to target dust continuum.

The data were calibrated, imaged and analysed using the standard ALMA pipeline and software package CASA \citep[version 4.5.3; ][]{mcmullin2007}.
The calibrated data were cleaned interactively using masks at source position and setting a threshold of 3$\times$r.m.s. noise level as measured on the dirty images. 
We adopted a Briggs weighting scheme \citep{briggs1995} with a \emph{robust} parameter of 0.2 and a channel width of 100 km s$^{-1}$ as the best trade-off between sensitivity and spatial resolution, resulting in a clean beam of $FWHM$=0.6$\arcsec$$\times$0.5$\arcsec$, with a position angle (P.A.) of 60$\degree$ in band 6 and of $FWHM$=1.0$\arcsec$$\times$0.7$\arcsec$, with a position angle (P.A.) of 80$\degree$ in band 3.

We used the cleaned data-cubes to produce continuum and line intensity maps (\emph{moment 0}; Fig. \ref{alma}) and to study the gas kinematics (Sec. \ref{sec:disk}).
ALMA 1.4mm continuum map (band 6) was obtained by averaging all the line-free channels in the data-cube over a total velocity range of $\sim$ 4000 km s$^{-1}$. The rms noise level is $\sigma_{1.4mm}$=0.03 mJy/beam. We do not have a significant continuum detection in band 3.
We also combined our band 3 continuum data with those from \citet{popping2017} taken from the ALMA archive, after taking into account the different angular resolutions. We do not find a significant continuum detection also in the combined map. From the combined image we quote a 5$\sigma$ upper limit on band 3 flux of $\sim$0.05 mJy, that is consistent with the flux intensity predicted by the SED (Fig. \ref{ivanplot}).

\emph{Moment 0} maps of the CO(3-2) and CO(6-5) lines shown in Fig. \ref{alma} were obtained by integrating the line channels over the velocity range between -1000 and 1000 km s$^{-1}$. In band 6 we first subtracted in the \emph{uv} plane the continuum with the task \texttt{uvcontsub}. The noise level measured in line-free channels is 0.15 and 0.10 mJy/beam, respectively in band 3 and 6.

We derived the source size and fluxes by fitting an elliptical Gaussian to the visibility data (task \texttt{uvmodelfit}), thus avoiding the uncertainty related to the cleaning parameters for these quantities.
We measure line fluxes $I_{\rm CO(3-2)}$=0.82$\pm$0.12 Jy km$^{-1}$, consistent with that reported by \citet{popping2017}, and $I_{\rm CO(6-5)}$=1.21$\pm$0.13 Jy km/s and 1.4 mm continuum flux $S_{\rm 1.4mm}$=378$\pm$65 $\mu$Jy.
Flux errors account for both measurement error and the 10$\%$ absolute flux accuracy due to the calibrator.\\
\indent While previous ALMA observations of GMASS 0953 \citep{popping2017} could not constrain the size of the molecular gas because of their lower angular resolution (2$\arcsec$), we marginally resolve the target in band 6.
For the CO(6-5) line we measure a deconvolved $FWHM$=0.18$\arcsec$$\pm$0.06$\arcsec$ (with an axis ratio of 1.0$^{+0.0}_{-0.3}$), that corresponds to a radius $r_{\rm CO}$=0.5$\times FWHM \sim$ 0.75$\pm$0.25 kpc, and an intrinsic size of the continuum emission of $FWHM$=0.30$\arcsec$$\pm$0.09$\arcsec$ ($r_{\rm 1.4mm}$=1.24$\pm$0.37 kpc), consistent with the CO(6-5) line within the errors.   

We have performed Monte Carlo simulations in order to test the reliability of \texttt{uvmodelfit} errors on the size of our target. In particular, we simulated the observation of a mock galaxy with the same best-fit properties of our target, both in line and in continuum. We then created 100 realizations of the background noise to match our observations and measured the properties of the simulated sources using \texttt{uvmodelfit}. In both cases (i.e. line and continuum) the peak and the sigma of the distribution of the measurements are perfectly consistent with the \texttt{uvmodelfit} best-fit size and errors.

The emission centroids in ALMA and HST images are co-spatial, after accounting for the known systematic $\lesssim$0.5$\arcsec$ shift in the NW direction between ALMA and HST positions in the CDFS \citep[e.g.][]{rujopakarn2016, dunlop2017, ginolfi2017}.
%
%
%
\subsection{ISM MODELING}\label{sec:gas}
GMASS 0953 had already been observed with ALMA by \citet{popping2017} in bands 3 and 4, targeting CO(3-2), CO(4-3), and [C I](1-0) lines.
%
%
%
    \begin{figure}
	\centering
	\includegraphics[trim=0mm 2mm 0mm 125mm, height=45mm, width=85mm, clip=true]{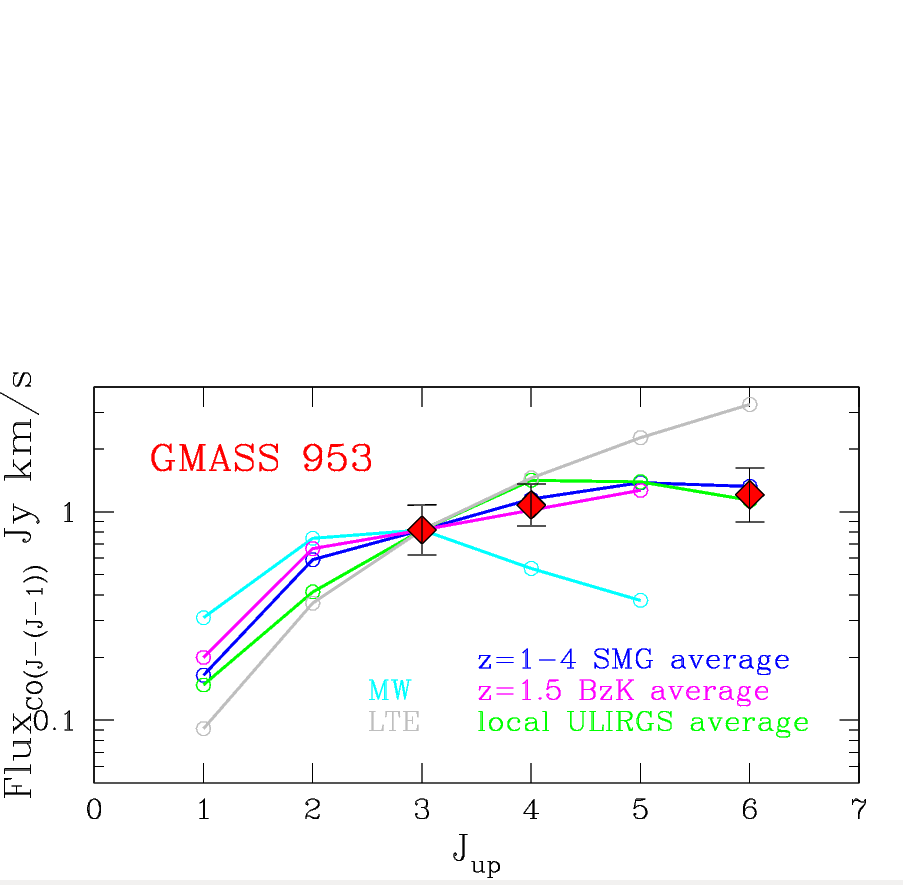}
	\caption{Observed CO SLED of GMASS 0953 (red diamonds).
We also plot, normalized to the J$_{\rm up}$=3 transition of our source, the expected scaling in the LTE approximation, the Milky Way \citep{fixsen1999}, and the average values from different classes of objects, namely SMGs \citep{bothwell2013}, BzK \citep{daddi2015}, and ULIRGs \citep{papadopoulos2012}. We added in quadrature a 10$\%$ flux accuracy uncertainty to the CO(4-3) line flux of GMASS 0953, that was not accounted for in \citet{popping2017} (Popping, private communication).}
	\label{sled}
	\end{figure}

They derived a CO(1-0) luminosity of $L^{\prime}$$_{\rm CO}$=2.1$\pm$0.2$\times$10$^{10}$ K km s$^{-1}$ pc$^{2}$ assuming that the lines are all in the Rayleigh-Jeans limit and in local thermodynamic equilibrium (LTE)\footnote[4]{The published value is actually a typo (Popping, private communication).}.
However, our new value of the CO(6-5) transition is different from what we would expect in the LTE approximation (Fig. \ref{sled}), therefore we adopt an empirical method for estimating the $L^{\prime}$$_{\rm CO}$ luminosity. 
The shape of the observed CO-SLED of GMASS 0953 shows a strong similarity with the average SLEDs of supposedly similar sources, namely BzK (mostly MS galaxies at $z$$\sim$2) and local ULIRGs (often hosting an AGN), normalized to the flux of the J$_{\rm up}$=3 transition of our target (Fig. \ref{sled}). 
Extrapolating the CO(1-0) transition from our observed CO(3-2) flux, assuming the average flux ratio from the appropriate literature SLEDs, we estimate a flux $I_{\rm CO(1-0)}$=0.17$\pm$0.03 Jy km s$^{-1}$. 
A similar value would be derived normalizing the SLEDs to the CO(6-5) flux of our target. 

From the CO(1-0) flux, following \citet{solomon1997} we derive $L^{\prime}$$_{\rm CO}$=(4.0$\pm$0.7)$\times$10$^{10}$ K km s$^{-1}$ pc$^{2}$, about twice the value quoted by \citet{popping2017}.
Assuming a CO-to-H$_{\rm 2}$ conversion factor $\alpha_{\rm CO}$=0.8 $M_{\odot}$/(K km s$^{-1}$ pc$^{2}$) we derive the gas mass: $M_{\rm H_{\rm 2}}$=(3.24$\pm$0.6)$\times$10$^{10}$$M_{\odot}$, that is in good agreement with the estimate derived from the [C I] emission line and a factor of $\sim$4 higher than that estimated from the dust mass \citep{popping2017}.
Our choice of $\alpha_{\rm CO}$ is motivated by the compactness of our source and its high SFR surface density \citep[Sec. \ref{sec:discussion}; see also][]{bolatto2013}.\\
\indent Concerning the ISM physical properties, \citet{popping2017} derived an estimate of the molecular gas density and the far-UV (6-13.6 eV) radiation field flux  from the comparison of the [C I]/CO(4-3) intensity ratio of GMASS 0953 to the outputs of single photo-dissociation region (PDR) models from \citet{kaufman1999, kaufman2006}. 
We did the same investigation with the code \emph{CLOUDY} v17.00 \citep{ferland2017}, adding our new observations. 
In particular, we run a grid of CLOUDY PDR models that span ranges in density and intensity of the UV radiation field that illuminates the cloud, assumed to be a 1-D gas slab, and we linearly scaled with the SFR the CLOUDY default Cosmic-ray Ionization Rate \citep[see][]{bisbas2015, vallini2017}.

\begin{figure}
	\centering
	\includegraphics[trim=0mm 0mm 0mm 0mm, height=45mm, width=85mm, clip=true]{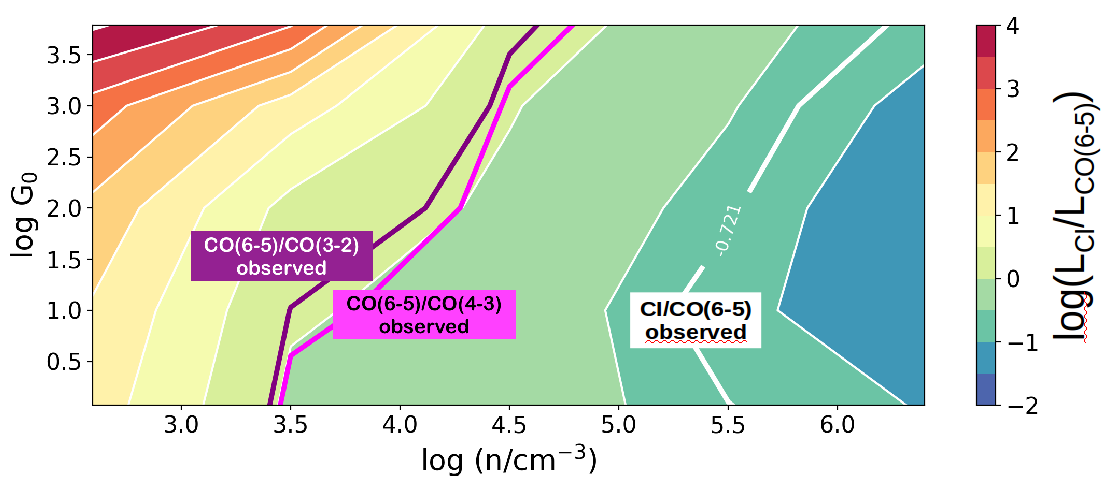}
	\caption{The [C I]/CO(6-5) luminosity ratio (expressed in erg s$^{-1}$ cm$^{-2}$) as a function of the $G_{0}$ and density \emph{n}.
The observed values of [C I]/CO(6-5), CO(6-5)/CO(4-3) and CO(6-5)/CO(3-2) are marked respectively by the thick white, magenta and purple contours.}
	\label{liviaplot}
	\end{figure}

In Fig. \ref{liviaplot} we show the predicted [C I]/CO(6-5) luminosity ratio as a function of $G_{0}$\footnote[5]{$G_{0}$ is the flux in the far-ultraviolet band (6-13.6eV) scaled to that in the solar neighborhood (~1.6$\times$10$^{-3}$ erg s$^{-1}$ cm$^{-2}$) \citep{habing1968}.} and density \emph{n}, highlighting the parameters space that give the observed [C I]/CO(6-5) (white) and CO(6-5)/CO(4-3) (magenta) luminosity ratios. 

It is evident that a single PDR with constant density and $G_{0}$ is not able to reproduce both luminosity ratios, because the two ratios do not trace the same parameters space. 
We argue that at least two components are needed to correctly model the observations, though a robust fit is not currently feasible because the degrees of freedom outnumber the data. 
Multiple phases are usually required to fit the ISM in local LIRGs and ULIRGs sources and in high-redshift galaxies \citep[e.g.][]{ward2003, carilli2010, danielson2011, daddi2015, pozzi2017, mingozzi2018}, consisting in a diffuse, lower-excitation component and a more concentrated, higher-excitation gas. 
Our measurement of the CO(6-5) transition points towards the existence of a very dense ISM component with $n\sim10^{5.5}$ cm$^{-3}$.

We also point out that GMASS 0953 is hosting a Compton-thick AGN and therefore higher-excitation emission could also be associated to a dense \emph{X-ray dominated region} (XDR), though higher-J CO lines would be needed to properly constrain its contribution \citep{meijerink2007, vanderwerf2010, pozzi2017, mingozzi2018}.
%
%
%
%
%
%
\section{Kinematics}\label{sec:disk}
The CO(3-2) and CO(6-5) lines have a $FWHM$ of 733$\pm$98 and 751$\pm$40 km s$^{-1}$ respectively, that are consistent with the FWHM of CO(3-2), CO(4-3) and [C I](1-0) reported by \citet{popping2017}.

\begin{figure}
	\centering
	\includegraphics[scale=0.27]{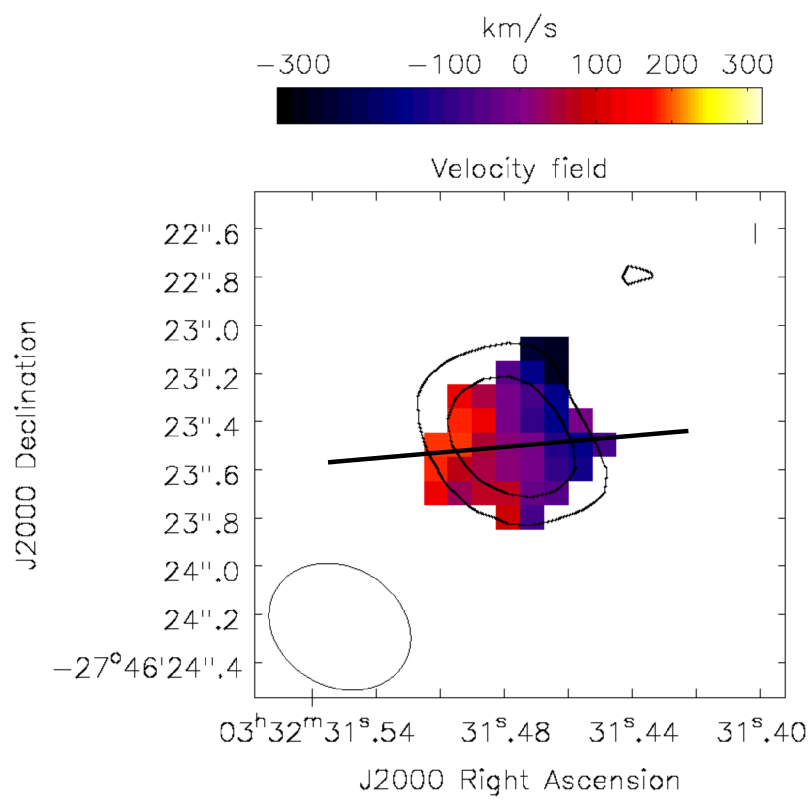}
	\caption{CO(6-5) velocity map with the continuum superimposed (black contours, 3$\sigma$ and 6$\sigma$, see Fig. \ref{alma}). The black line shows the direction of the major axis.}
	\label{velmap}
	\end{figure}

\begin{figure}
	\centering
	\includegraphics[scale=0.24]{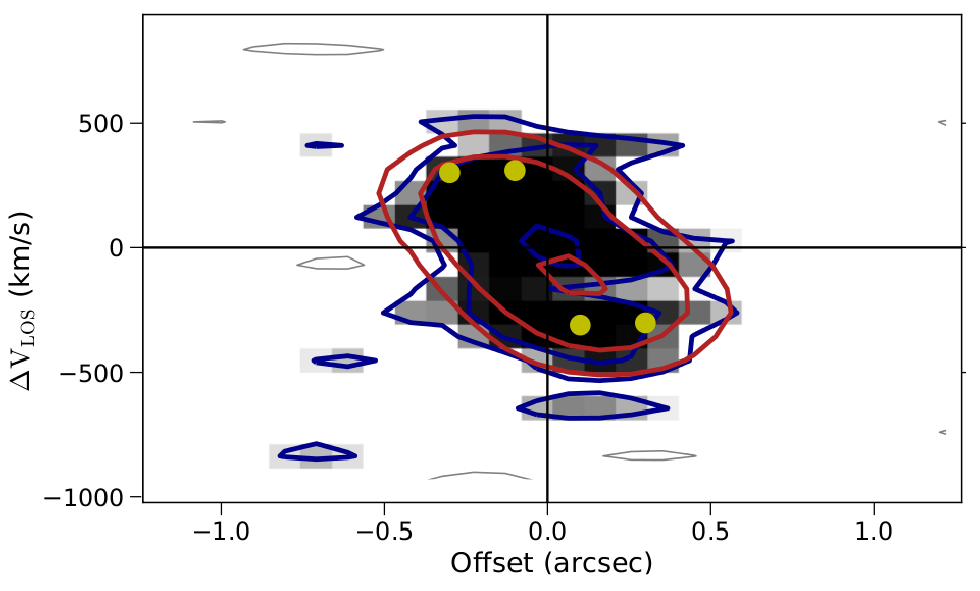}
	\caption{CO(6-5) PV diagram extracted along the major axis at $\phi$=95$\degree$, assuming an inclination of 75$\degree$. 
The the iso-density contours (2.5, 5, 10$\sigma$) of the galaxy and \texttt{BAROLO} best-fit model are shown in blue and red, respectively. 
The yellow points mark the rotation curve.}
	\label{pv}
	\end{figure}

We show the CO(6-5) velocity map in Fig. \ref{velmap} and the position-velocity (PV) diagram extracted along the major axis at a P.A. of 95$\degree$ in Fig. \ref{pv}. A velocity gradient is clearly detected. 
A merger system in a coalescence phase observed at a favourable orientation could in principle originate such a gradient. 
However, this picture seems unlikely based on the absence of two distinct nuclei in the core of the high-resolution HST/ACS image (Fig. \ref{hst}).
Alternatively, the PV diagram could be the signature of a rotating disk of molecular gas.
Possible evidence of a rotating disk of ionized gas in GMASS 0953 from the study of the H$\alpha$ and [OIII]$\lambda$5007 emission lines kinematics has been also reported \citep[][Loiacono et al. in prep.]{wisnioski2017}.

Under the assumption of a rotating disk we investigate the kinematic properties of the dense molecular gas traced by the CO(6-5) line with \texttt{$^{3D}$BAROLO} \citep{diteodoro2015}, a tool for fitting 3D tilted-ring models to emission-line datacubes that takes into account the effect of beam smearing.
We assumed a disk model with two rings and a ring width of 0.2$\arcsec$, that is approximately half the clean beam size of the datacube.
We fix the P.A. at 95$\degree$, that is the value that maximizes the spatial extention of the galaxy in the PV diagram. 
This value is consistent with both the photometric and kinematic H$\alpha$ PAs as determined by HST imaging and KMOS data \citep{vanderwel2014, wisnioski2017}. 
Then we run \texttt{BAROLO} leaving as free parameters the rotation velocity ($V_{rot}$) and the intrinsic velocity dispersion ($\sigma$) for different values of the inclination. 
%
%
%

\begin{figure}
	\centering
	\includegraphics[scale=0.27, trim=0mm 0mm 0mm 70mm, clip=true]{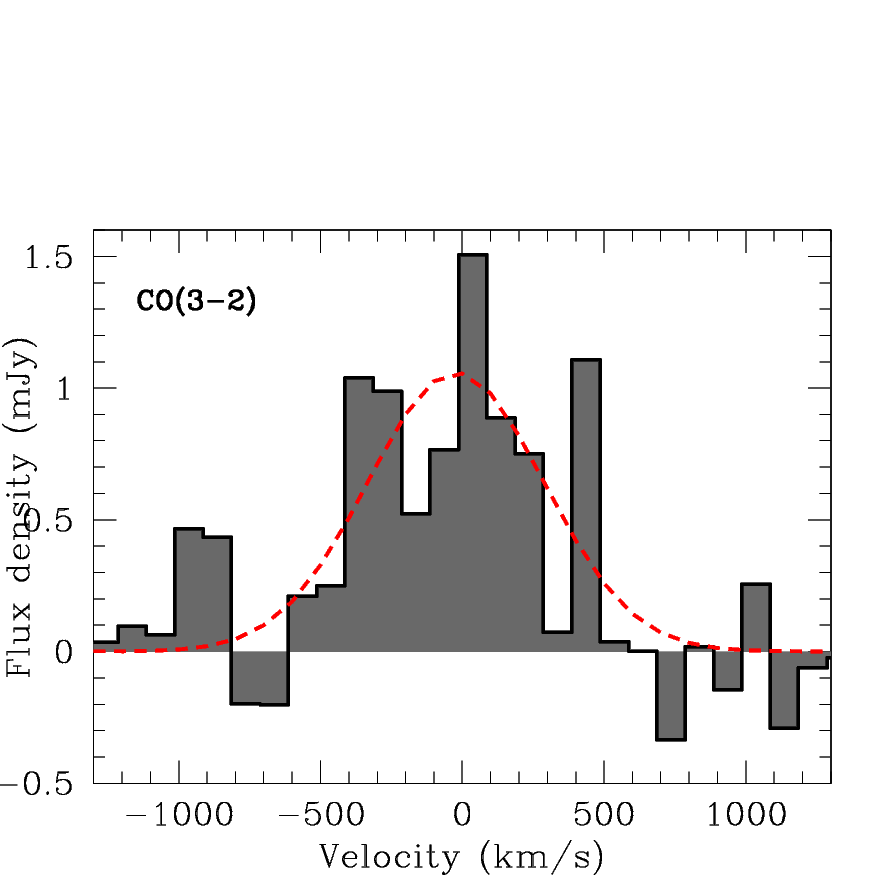} 
	\includegraphics[scale=0.27, trim=0mm 0mm 0mm 70mm, clip=true]{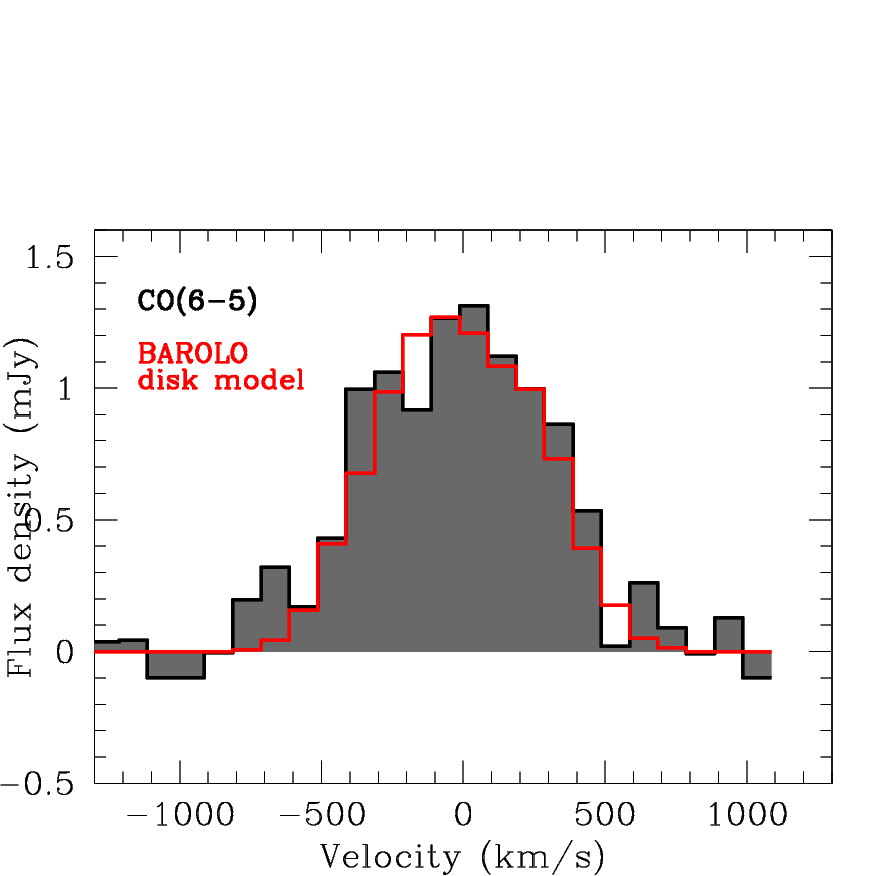} 
	\caption{Velocity-integrated flux densities of the CO(3-2) (top) and CO(6-5) (bottom). 
Both spectra were extrated from the respective data-cubes in the region delimited by the 3$\sigma$ contour in the \emph{moment 0} map (see Fig. \ref{alma}).
The red dashed line in the top plot marks the gaussian fit to the density profile. 
The red continuous line in the bottom plot is the disk model extracted from the 3D-BAROLO model-cube in same region as the source spectrum.}
	\label{spec}
	\end{figure}

We derive a fiducial interval for the inclination (i.e. the range of values for which the model does not change significantly, as estimated from the residuals maps) between 60$\degree$ and 90$\degree$ and a best-fit $V_{rot}$=320$^{+92}_{-53}$ km s$^{-1}$, where the error includes both the formal error from the fit and the uncertainty from the variation of the inclination in our fiducial range. 
From our simulations we also conclude that the model is quite insensitive to large variations in $\sigma$ due to the large channel width and estimate an upper limit of $\sigma$=140 km/s.
The best-fit model normalized to the azimuthally-averaged flux in each ring is shown in red contours in Fig. \ref{pv}. 
We also show the 1D spectrum extracted from the model, along with the CO(3-2) spectrum, in Fig. \ref{spec}.

We note that the intrisic H$\alpha$ rotation curve presented in Fig. 5 of \citet{wisnioski2017} suggests an intrinsic velocity of $\sim$200 km s$^{-1}$ on nuclear scale, broadly consistent with our results, and a lower velocity of $\sim$100 km s$^{-1}$ at larger radii. Though it is difficult to make a direct comparison with the results by \citet{wisnioski2017} because of the different approaches to deal with beam-smearing effects, we speculate that the combination of the two results might suggest a declining rotation curve of the inner regions of the gas disk in GMASS 0953 as observed in massive early-type galaxies both locally and at high-redshift \citep{noordermeer2007, genzel2017}.

We note a 2.5$\sigma$ flux excess with respect to the disk model with an offset of $\Delta v\sim$ -700 km s$^{-1}$ with respect to the line peak, also visible in the PV diagram. 
The velocity offset is consistent with the signatures of AGN-driven large-scale outflows in the neutral and ionized gas phases, namely the blueshift of rest-frame UV ISM absorption lines and the offset of a broad component detected in the [OIII]$\lambda$5007 emission line \citep[][Loiacono et al. in prep]{cimatti2013}.
\citet{popping2017} report that they do not find any signature of outflow in the flux density profile of the CO(4-3) line, that is detected at a similar significance level as the CO(6-5) line in this work. The lack of a flux excess in the lower-J observed transitions could indicate that the excitation ratio between the CO(6-5) and lower-J transitions is higher in the possible outflow than in the rest of molecular gas in the host galaxy \citep[e.g.][]{dasyra2016, richings2017}.
%
%
%
%
%
%
%
%
%
\section{Discussion}\label{sec:discussion}
We have presented ALMA observations of GMASS 0953, an heavily obscured AGN host at z$\sim$2.226.
The M$_{H_{2}}$ derived in Sec. \ref{sec:gas} returns a gas fraction $M_{\rm H_{\rm 2}}$/($M_{\rm H_{\rm 2}}$+$M_{\star}$)=0.2 and a gas depletion time scale $\tau_{\rm depl}$=$M_{\rm H_{\rm2}}$/$SFR$$\sim$150 Myr.
As pointed out by \citet{popping2017} this value of $\tau_{\rm depl}$ is much shorter than in more extended MS galaxies at the same redshift \citep{sargent2014, scoville2017, tacconi2017}, but consistent with the values measured in off-MS galaxies, other cSFGs, and a few galaxies hosting an obscured AGN \citep[e.g.][]{polletta2011, brusa2015b, barro2016, spilker2016, tadaki2017b}.
We find evidence for a multi-phase ISM in our galaxy and estimate the density of the higher-excitation gas probed by the observed CO(6-5) line: $n\sim10^{5.5}$ cm$^{-3}$. 

We measure a very compact radius ($\sim$1 kpc) for both the molecular gas and the dust emission, $\sim$2 times smaller than the stellar distribution. 
We derive a gas mass surface density of $\Sigma_{M_{\rm H_{\rm 2}}}$=0.5$\times$$M_{\rm H_{\rm 2}}$/$\pi$($r_{\rm CO}$)$^{2}$$\sim$9000 $M_{\odot}$ pc$^{-2}$. 
This value is similar to the typical stellar mass surface density of quiescent galaxies of similar stellar mass at the same redshift \citep{barro2015}.
Considering that the SFR surface density at the radius of the dust continuum is $\Sigma_{SFR}$=0.5$\times$$SFR$/$\pi$($r_{\rm 1.4mm}$)$^{2}$$\sim$22 $M_{\odot}$ yr$^{-1}$ kpc$^{-2}$ GMASS 0953 would lie at the high star-formation and gas-density end of the Kennicutt-Schmidt relation, consistent with local ULIRGs \citep[e.g.][]{genzel2010}.\\
\indent From the afore-mentioned results we conclude that GMASS 0953, though formally lying on the MS of SFGs, has a lower gas content than MS galaxies with the same stellar mass and is consuming it much more rapidly in a very compact core \citep{elbaz2017}.
On short timescales this galaxy will likely exhaust its gas reservoirs and become a cQG.
This scenario, consistent with previous analysis of ISM properties and optical emission lines kinematics \citep{vandokkum2015, popping2017, wisnioski2017}, finds a further confirmation from our direct measurement of the extremely compact size of the star-forming region and the molecular gas of GMASS 0953 \citep{gilli2014, barro2016, tadaki2017a, tadaki2017b, brusa2018}.
With our data we are unable to discriminate between the different mechanisms that could have originated such a compact core, i.e. a merger occurred in the past, \citep[e.g.][]{tacconi2008, wellons2015}, disk instabilities \citep[e.g.][]{dekel2014, ceverino2015, zolotov2015}, or in-situ secular processes \citep[e.g.][]{wellons2015, vandokkum2015}, though recent studies tend to favour dissipative formation mechanisms to explain the smaller size of the nuclear region of intense star formation with respect to the stellar distribution \citep{barro2016, tadaki2017a}. 

AGN activity is also advocated as a quenching mechanism for cSFGs in addition to the gas consumption provided by the strong star-formation activity \citep{barro2013}, both likely triggered by the same mechanism that led to the formation of the compact core \citep{kocevski2017}.
In particular, because of its compactness and the presence of a luminous, obscured AGN, GMASS 0953 is consistent with the 'quasar mode' postulated by \citet{hopkins2006b} where the AGN quenches the star formation within the host galaxy through feedback mechanisms \citep[see also][and Lapi et al. in prep.]{rangel2014}, e.g. fast large-scale outflows as those that have been observed in different gas phases of GMASS 0953 \citep[][Loiacono et al. in prep.]{cimatti2013, forsterschreiber2014, genzel2014}, tentatively including the molecular one.

GMASS 0953 is also one of the first cases in which, thanks to the quality of the data, we are able to measure the rapid rotation ($V_{rot}$=320$^{+92}_{-53}$ km s$^{-1}$) of the molecular gas disk in the core \citep{tadaki2017b, barro2017, brusa2018}, predicted by some simulations before the gas is completely depleted \citep{shi2017}, though it is not yet clear if this is a common feature in all cQGs progenitors \citep[e.g.][]{spilker2016}. 
The observation of stellar rotation in cQGs \citep{newman2015, toft2017} could indicate that cSFGs cores might retain their rotation after the quenching processes. 

In conclusion, in this work we have highlighted the importance of spatially-resolved ALMA observations for the study of a prototypical progenitor of cQGs, likely caught in the act of quenching through the combined action of efficient compact nuclear star-formation activity and AGN feedback.
%
%
%
\section{Acknowledgements}
This paper makes use of the following ALMA data: ADS/JAO.ALMA$\#$2015.1.01379.S (PI: Cassata); $\#$2015.1.00228.S (PI: Popping). 
ALMA is a partnership of ESO (representing its member states), NSF (USA) and NINS (Japan), together with NRC (Canada), NSC and ASIAA (Taiwan), and KASI (Republic of Korea), in cooperation with the Republic of Chile. The Joint ALMA Observatory is operated by ESO, AUI/NRAO and NAOJ.
We acknowledge extensive support in data reduction and analysis from the ALMA Regional Centre in Bologna.
MT gratefully thanks E. Di Teodoro for his support with the \texttt{$^{3D}$BAROLO} code, P. Popesso for her warm hospitality in Munich during the writing of this paper, L. Pantoni for providing the dust mass and temperature of GMASS 0953, G. Popping, R. Decarli, R. Paladino and A. Lapi for useful discussions.
PC acknowledges support from CONICYT through the project FONDECYT regular 1150216. 
MB acknowledges support from the FP7 Career Integration Grant "eEASy" (CIG 321913). 
FP, CG, AR, LP, GR acknowledge funding from the INAF PRIN-SKA 2017 program 1.05.01.88.04.
EI acknowledges partial support from FONDECYT through grant N$^\circ$\,1171710.
The authors thank the anonymous referee for constructive comments that helped to improve the presentation of the results.

%
%
%

\bibliographystyle{mnras}
\bibliography{references} 
\bsp	
\label{lastpage}
\end{document}